\documentclass{pnastwo}

\usepackage[pdftex]{graphicx}
\usepackage[sort&compress,numbers]{natbib}

\usepackage{amssymb,amsfonts,amsmath}

\url{www.pnas.org/cgi/doi/10.1073/pnas.0709640104}
\copyrightyear{2008}
\issuedate{Issue Date}
\volume{Volume}
\issuenumber{Issue Number}

\begin{document}

\title{Separating astrophysical sources from indirect dark matter signals}

\author{Jennifer~M.~Siegal-Gaskins\affil{1}{Einstein Fellow}\affil{2}{California Institute 
of Technology, 1200 E. California Blvd., Pasadena, CA 91125}}

\contributor{Submitted to Proceedings of the National Academy of Sciences
of the United States of America}

\maketitle

\begin{article}
\begin{abstract} 
Indirect searches for products of dark matter annihilation and decay face the challenge of identifying an uncertain and subdominant signal in the presence of uncertain backgrounds. Two valuable approaches to this problem are (1) using analysis methods which take advantage of different features in the energy spectrum and angular distribution of the signal and backgrounds, and (2) more accurate characterization of backgrounds, which allows for more robust identification of possible signals.  
These two approaches are complementary and can be significantly strengthened when used together.
I review the status of indirect searches with gamma rays using two promising targets, the Inner Galaxy and the Isotropic Gamma-Ray Background.  For both targets, uncertainties in the properties of backgrounds is a major limitation to the sensitivity of indirect searches.  I then highlight approaches which can enhance the sensitivity of indirect searches using these targets.
\end{abstract}


\dropcap{I}ndirect dark matter detection offers a promising approach to detecting dark matter in an astrophysical context, and may provide a means of mapping its spatial distribution and constraining its particle nature.  Current searches for signatures of dark matter annihilation or decay using gamma rays have made substantial progress in recent years~\cite{Ackermann:2011wa,GeringerSameth:2011iw,Hooper:2011ti,Ackermann:2012rg,Abazajian:2012pn,Hooper:2012sr,Gordon:2013vta}, with sensitivity to dark matter annihilation beginning to reach the favored regions of parameter space for thermal relic weakly-interacting massive particle (WIMP) dark matter~\cite{Jungman:1995df,Steigman:2012nb}.  

The most robust constraints on indirect dark matter signals are obtained by placing limits on the total amplitude of a dark matter signal without attempting any background modeling.  For a few targets, such as Milky Way dwarf spheroidal galaxies~\cite{Ackermann:2011wa,GeringerSameth:2011iw}, the expected backgrounds are small, and thus the sensitivity is not significantly weakened by neglecting to model the backgrounds.  However, for many favorable indirect search targets in gamma rays, including the Inner Galaxy and the Isotropic Gamma-Ray Background (IGRB), the guaranteed backgrounds are substantial and currently not strongly constrained.  As a result, dark matter searches in these regions could be made considerably more sensitive by improved knowledge of backgrounds.  

While most indirect searches to date have resulted in limits being placed, in some cases intriguing hints of a possible signal have been uncovered. Several groups have reported an excess in GeV gamma-ray emission originating from the Galactic Center~\cite{Hooper:2010mq,Hooper:2011ti,Abazajian:2012pn,Gordon:2013vta,Abazajian:2014fta,Daylan:2014rsa} and Inner Galaxy~\cite{Hooper:2013rwa,Daylan:2014rsa}; here I use Galactic Center to refer to  the region within a few degrees of the dynamical center of the Galaxy, and Inner Galaxy to refer to the region extending a few tens of degrees from the Galactic Center.  However, in all of these studies the putative dark matter signal was detected using an approach that required some form of background modeling.  Due to uncertainties in the emission from non-dark-matter sources in this region, the interpretation of these possible detections of excess emission over a background model as a dark matter signal has been challenged, see, e.g.~\cite{Boyarsky:2010dr,Abazajian:2010zy}; see also~\cite{Hooper:2013nhl}.

Indirect searches focusing on the IGRB have generated interesting constraints on dark matter properties~\cite{Abdo:2010dk,Abazajian:2010zb}, however these limits depend strongly on the expected contributions to the IGRB from other gamma-ray source populations.  The IGRB is guaranteed to contain contributions from undetected members of the many gamma-ray source classes that already have confirmed members~\cite{Pavlidou:2007su}.  The recent sharp increase in the number of detected and characterized gamma-ray sources by the \emph{Fermi} Large Area Telescope (LAT) and Imaging Atmospheric Cherenkov Telescopes (IACTs) such as H.E.S.S., VERITAS, and MAGIC make improved modeling of IGRB contributions possible via population studies of detected sources.  The fact that the observed spectral shape of the IGRB does not exhibit any distinct features, combined with the expected levels of contributions from confirmed source classes, strongly suggests that any dark matter signal will be subdominant.  A more accurate determination of the contributions of non-dark-matter sources to the IGRB could thus have a major impact on indirect searches.

Over the last $\sim 5$~years the \emph{Fermi} LAT has accumulated high-quality all-sky data at energies from $\sim 100$~MeV to $\gtrsim 500$~GeV, and continues to collect data and to refine the characterization of the instrument~\cite{Ackermann:2012kna,Atwood:2013rka}.  At the same time, ground-based IACTs have opened a new window on the TeV sky, and the upcoming Cherenkov Telescope Array (CTA) is expected to deliver unprecedented sensitivity and angular resolution at energies above $\sim 100$~GeV, with a larger field-of-view and effective area than currently-operating IACTs (e.g.,~\cite{Funk:2012ca}).  It is thus timely to reevaluate indirect dark matter search strategies to determine the most effective approaches for current and upcoming data sets and to identify ways to fully take advantage of the capabilities of future instruments.

In this paper I focus on two favorable targets of indirect searches, the Inner Galaxy and the IGRB, and discuss approaches to improving sensitivity to dark matter signals.  Spectral information (including multi-wavelength studies spanning a broad range of energies) and spatial information are key discriminants that have been leveraged in past searches, and should continue to play a role in future searches.  In this work I also emphasize the importance of an improved understanding of sources of background and impostor signals, and the potential of such improvements to enable more sensitive dark matter searches.  Although here I discuss techniques for extending the reach of indirect searches with photons, indirect searches using charged particles and neutrinos also provide important, complementary information (see~\cite{Rott:2012gh} and references therein).  Making use of such complementary results is essential to maximize returns from the data by using all available information, to alleviate issues with systematics associated with a single experiment, and to help to efficiently direct future efforts.

\section{The Inner Galaxy}

The density of dark matter increases rapidly toward the Galactic Center, making the Inner Galaxy one of the strongest sources of high-energy emission from dark matter in the sky, and one of the most optimistic targets of indirect searches.  However, the Galactic Center is one of the most complicated regions of the sky in gamma rays, as it is home to a wide variety of astrophysical sources, including the supermassive black hole Sgr A$^{\star}$, a supernova remnant, and a giant molecular cloud complex.

H.E.S.S. observations of the inner few degrees around the Galactic Center at energies greater than 380 GeV revealed a point-like source of high-energy emission at the dynamical center of the Milky Way, as well as evidence for emission extending several degrees that is strongly spatially correlated with molecular clouds~\cite{Aharonian:2006au}, see also~\cite{Tsuboi:1999gc}.  The spectrum of the central point source is compatible with a power law over the energy range analyzed and thus shows no clear evidence for a dark matter origin~\cite{Aharonian:2006wh}, which would instead imply a cut-off in the spectrum at the energy corresponding to the dark matter particle mass for an annihilation signal (half of the dark matter particle mass in the case of decay).

Several groups have analyzed \emph{Fermi} LAT observations of the Galactic Center, however no consensus on their interpretation has emerged.  While some groups have claimed a significant extended excess at the Galactic Center at GeV energies, consistent with dark matter expectations and incompatible with models of detected point sources and diffuse emission~\cite{Hooper:2011ti,Abazajian:2012pn,Abazajian:2014fta,Daylan:2014rsa}, others have questioned the analysis approach that revealed the excess~\cite{Boyarsky:2010dr} or offered alternative explanations for the claimed signals~\cite{Abazajian:2010zy,YusefZadeh:2012nh}.   

In all of these cases of claimed detections, the excess emission was identified by performing some type of modeling of the astrophysical emission in the region, and finding that there was additional emission beyond what could be accommodated by the model.  Substantial uncertainties exist in the models of diffuse gamma-ray emission from the Galactic Center due to limited knowledge of cosmic-ray sources and the interstellar gas and radiation fields~\cite{FermiLAT:2012aa}.  
In principle one could use the fact that the gamma-ray spectra resulting from dark matter annihilation and decay are expected to differ somewhat from that of the diffuse background emission, in order to identify a signal of dark matter origin.  However due to the complexity of the region and the fact that any diffuse model of the Galactic Center includes several components with distinct spectra that are summed to account for the total emission, it is difficult to dismiss the possibility that excesses at certain energies are simply artifacts resulting from inaccuracies in or incompleteness of the diffuse model.

Furthermore, undetected members of Galactic source populations also contribute to the measured diffuse emission from the Galactic Center region~\cite{Strong:2006hf}.  Of special concern for a dark matter study is the emission from unresolved pulsars, which can exhibit a similar gamma-ray energy spectrum to that associated with some dark matter models.  In particular, gamma-ray millisecond pulsars (MSPs) have frequently been cited as a possible impostor signal not only because of their energy spectrum but also because their expected angular distribution in the Inner Galaxy could mimic that of a dark matter signal (e.g.,~\cite{Abazajian:2010zy,Gordon:2013vta}).  MSPs are pulsars which have been ``spun-up'' as a result of accretion from a binary companion.  MSPs can be as luminous as ordinary pulsars, but have shorter periods, evolve more slowly, and live much longer.  MSPs typically have much lower velocities than ordinary pulsars, which would allow them to remain near the Galactic Center if formed there, while their longer lifetimes could enable MSPs formed throughout the Galaxy to slowly migrate away from the Galactic plane; ordinary pulsars would likely become faint before traveling very far from where they were formed.  Consequently, similar to the emission from dark matter annihilation, the MSP spatial distribution may be concentrated near the Galactic Center but also extend significantly away from the plane.  Based on the typical observed gamma-ray fluxes of MSPs, it has been suggested that a population of $\sim 1000$ MSPs in the Galactic Center could explain the GeV excess~\cite{Gordon:2013vta}, although it has not been established that such a population of MSPs exists in the Galactic Center.

\begin{figure}[t]
\centering
\includegraphics[width=.45\textwidth]{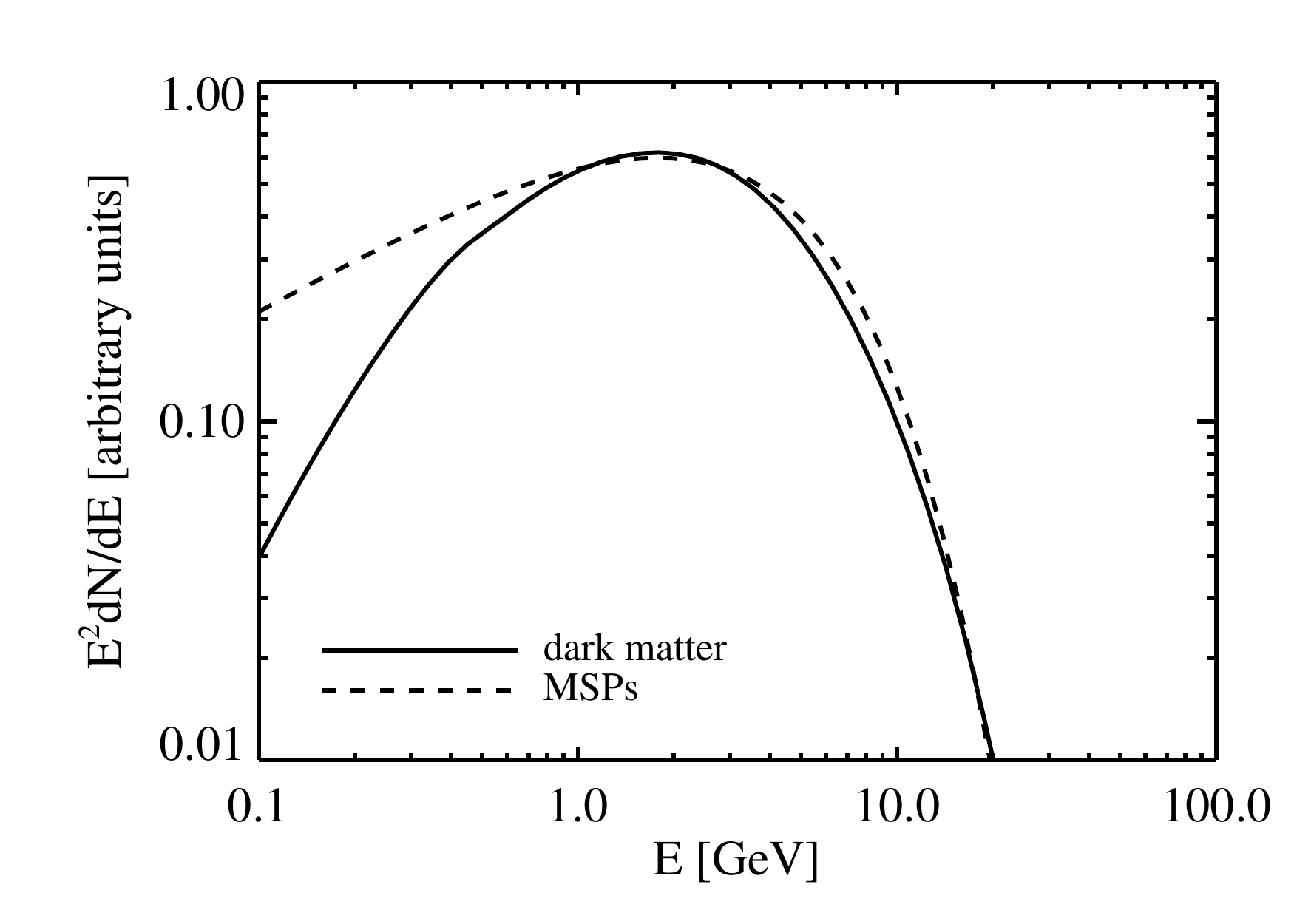}
\caption{Spectrum of gamma rays from annihilation of 35~GeV dark matter particles to $b\bar{b}$ (solid curve), which represents a good fit to the Galactic Center GeV excess, compared with the best-fit spectrum for 37 gamma-ray MSPs detected by the \emph{Fermi} LAT (dashed curve) as determined in~\cite{Hooper:2013nhl}.  The MSP spectrum is not sufficiently hard to fit the low-energy part of the spectrum of the Inner Galaxy excess.\label{fig:IGpulsars}}
\end{figure}

Recently an excess with a similar spectrum to that of the Galactic Center GeV excess has been claimed in \emph{Fermi} data at higher Galactic latitudes~\cite{Hooper:2013rwa,Daylan:2014rsa}, with an angular distribution and energy spectrum consistent with expectations for annihilation of light WIMP dark matter.  Although the detection of this Inner Galaxy excess is still subject to uncertainties in the diffuse emission model and the contribution of unresolved sources, at higher latitudes the uncertainties in the diffuse model may be somewhat reduced, making this a particularly intriguing claim.  It remains unclear whether unresolved MSPs could explain the Galactic Center GeV excess, however it appears that the energy spectrum of typical MSPs is mildly inconsistent with that of the Inner Galaxy excess (Fig.~\ref{fig:IGpulsars}), and moreover commonly-adopted models of the luminosity and spatial distribution of the Galactic MSP population cannot account for the amplitude of the excess without severely overproducing the number of bright gamma-ray MSPs~\cite{Hooper:2013nhl}.   Even though it may be difficult to account for the entirety of the claimed high-latitude excess with unresolved MSPs, it is likely that MSPs contribute at some level, and accounting for their contribution could lead to different spectral properties or a modified angular distribution of the remaining excess emission, which may challenge dark matter interpretations.

Other possible issues for a dark matter interpretation of the Galactic Center excess have been identified when jointly considering the H.E.S.S. and \emph{Fermi} data.   A component of the central point source spectrum can be well-modeled in both data sets by a single power law spanning several decades in energy, but it is challenging to model both the GeV and TeV deviations from a power law with a single dark matter model~\cite{Belikov:2012ty}.  In addition, while variability is seen in radio (e.g.,~\cite{YusefZadeh:2010wz}) and X-ray (e.g.,~\cite{Aschenbach:2004fx}) observations of Sgr A$^{\star}$, no variability has been seen in gamma rays~\cite{Aharonian:2009zk}, suggesting that the source of the gamma-ray emission differs from that of the lower-energy signals.  The situation is further complicated by the differing angular resolution of the H.E.S.S and \emph{Fermi} data sets, making it difficult to compare the emission originating from the same regions of the sky.  However, in \cite{Linden:2012bp} the authors argue that the improved angular resolution of CTA will be sufficient to distinguish between a gamma-ray point source and a hadronic extended emission scenario at the Galactic Center.

Multi-wavelength observations offer an important means of testing dark matter interpretations (e.g.,~\cite{Regis:2008ij,Linden:2010eu,Hooper:2010im}) and characterizing astrophysical backgrounds and impostor signals in the Inner Galaxy (e.g.,~\cite{YusefZadeh:2012nh,Wang:2013dqq}), thereby enabling more sensitive indirect searches.  Cosmic rays produced by dark matter annihilation and decay as well as by astrophysical sources generate secondary emission spanning radio to gamma-ray energies due to interactions with the interstellar gas, photon fields, and magnetic fields.  Studies using lower-energy observations of the Galactic Center have generated important constraints on a variety of dark matter models (e.g.,~\cite{Bertone:2008xr,Crocker:2010gy,Laha:2012fg}).  Upcoming observations at a variety of wavelengths, e.g., those from LOFAR, NuSTAR, and CTA, in addition to recent data from \emph{Planck}, have the potential to significantly improve our understanding of astrophysical sources of emission and the environment of the Inner Galaxy, and to narrow the parameter space of viable dark matter models.

\section{The IGRB}

The IGRB is a faint all-sky glow at gamma-ray energies that has been detected and characterized by several instruments, most recently by the \emph{Fermi} LAT\@.  It is thought to represent the collective emission from both extragalactic and some Galactic gamma-ray sources which are not detected individually.
Note that the term IGRB is used to refer to emission which appears to be statistically isotropic, i.e., having no preferred direction; it does not mean that the emission is smooth and of equal intensity in all directions.  Although the intensity of the IGRB is in fact observed to be fairly uniform on large angular scales, it exhibits anisotropies on small angular scales which have been used to study its origin, as discussed below.

The vast majority of emission from dark matter annihilation and decay in structures throughout the Universe, as well as in our own Galaxy, is unlikely to be resolved into individual sources by current or upcoming instruments.  Consequently, the collective emission from these processes will contribute to the measured diffuse emission.   The IGRB is thus an important target for indirect dark matter searches.

In recent years our understanding of the contribution of several astrophysical source classes to the IGRB has improved greatly, while at the same time the question of the IGRB composition has become more complicated as it has become clear that many different source populations, both extragalactic and Galactic, contribute at some level.  
The source classes currently thought to provide the most important contributions are blazars~\cite{Collaboration:2010gqa,Abazajian:2010zb,Ajello:2011zi}, starforming galaxies~\cite{Ackermann:2012vca}, radio galaxies~\cite{Inoue:2011bm}, misaligned AGN~\cite{DiMauro:2013xta}, and Galactic millisecond pulsars~\cite{FaucherGiguere:2009df}.

\subsection{The IGRB intensity spectrum}

\begin{figure*}[t]
\centering
\includegraphics[width=.6\textwidth]{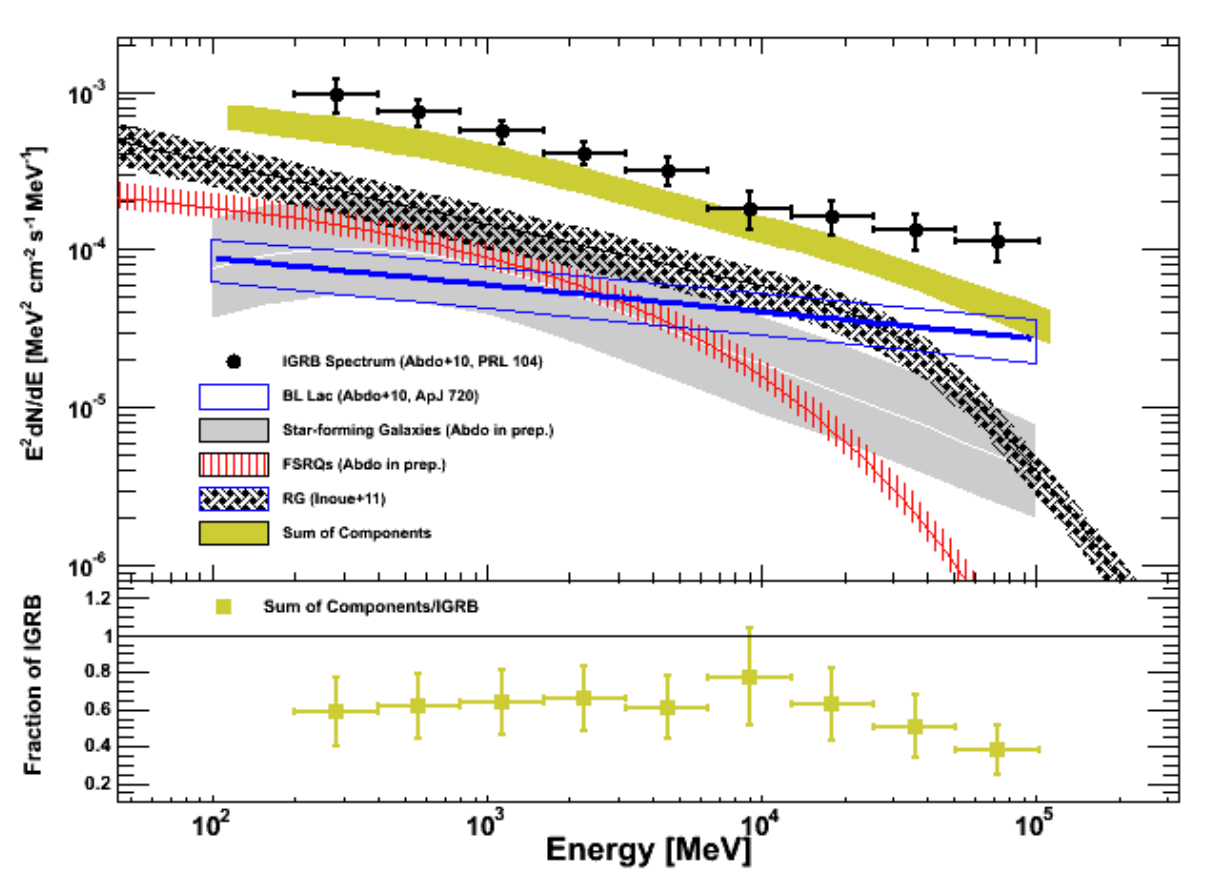}
\caption{Intensity spectrum of the IGRB~\cite{Abdo:2010nz}.  Predictions for contributions of selected gamma-ray source classes~\cite{Collaboration:2010gqa,Ackermann:2012vca,Ajello:2011zi,Inoue:2011bm} are also shown, with uncertainties.  The sum of the components shown falls short of the total measured emission at most energies, with large uncertainties at all energies.  From~\cite{Ajello:2013aps}. 
\label{fig:IGRBspectra}}
\end{figure*}

The intensity spectrum of the IGRB as measured by the \emph{Fermi} LAT~\cite{Abdo:2010nz} along with the expected contributions to the IGRB from several confirmed gamma-ray source populations are shown in Fig.~\ref{fig:IGRBspectra}.  It is notable both that the sum of the contributions shown in Fig.~\ref{fig:IGRBspectra} is insufficient to account for the measured IGRB intensity, and that the uncertainty on the contribution of each individual source class is large.  This uncertainty in the composition of the IGRB intensity poses a challenge for performing highly sensitive indirect dark matter searches since within the uncertainty there remains a great deal of room for a dark matter component in the IGRB\@.  On the other hand, reducing the uncertainty on the energy-dependent level of emission contributed by non-exotic sources has the potential to significantly increase the reach of indirect dark matter searches by more precisely pinning down the energy spectrum and amplitude of the emission available to be attributed to dark matter signals.

The fact that new individual gamma-ray sources will be resolved with future observations has two important implications for improving dark matter searches in the IGRB\@.  First, as more sources are resolved, the intensity of the IGRB will decrease, reducing the emission available to possible dark matter signals~\cite{Abazajian:2010pc}.  Second, as the number of detected and characterized sources increases, population models of sources contributing to the IGRB will improve, enabling more robust estimates of the collective contribution of known sources to the IGRB intensity.

\subsection{Anisotropies in the IGRB}
In addition to the spectral information encoded in its total intensity spectrum, the IGRB also contains spatial information which can be used to reveal its components, in the form of small-angular-scale fluctuations.  These anisotropies in the IGRB provide a statistical measure of the properties of contributing source populations (the anisotropy encodes the flux distribution and clustering properties of the sources).
Several studies have predicted the anisotropy arising from specific source classes that contribute to the IGRB, including blazars~\cite{Ando:2006cr,Miniati:2007ke,Cuoco:2012yf,Harding:2012gk}, starforming galaxies~\cite{Ando:2009nk}, Galactic millisecond pulsars~\cite{SiegalGaskins:2010mp}, and Galactic and extragalactic dark matter annihilation and decay~\cite{Ando:2005xg,Ando:2006cr,SiegalGaskins:2008ge,Fornasa:2009qh,Ando:2009fp,Fornasa:2012gu}.

The first measurement of small-scale anisotropy in the IGRB was recently made with the \emph{Fermi} LAT~\cite{Ackermann:2012uf}.  Angular power was measured in four energy bins spanning 1--50~GeV, with significant ($> 3\sigma$) detections in the three bins covering 1--10~GeV.  This measurement has lead to new constraints on source populations by requiring consistency between the predicted and measured anisotropy.  In particular, strong constraints have been obtained on blazar population models, severely limiting their contribution to the IGRB intensity to less than $\sim 20$\%~\cite{Cuoco:2012yf,Harding:2012gk} and excluding certain proposed models which explain a larger fraction of the IGRB intensity.  In addition, independent constraints on dark matter models have been derived from the anisotropy measurement~\cite{Gomez-Vargas:2013cna,Ando:2013ff}.

\begin{figure*}[t]
\centering
\includegraphics[width=.45\textwidth]{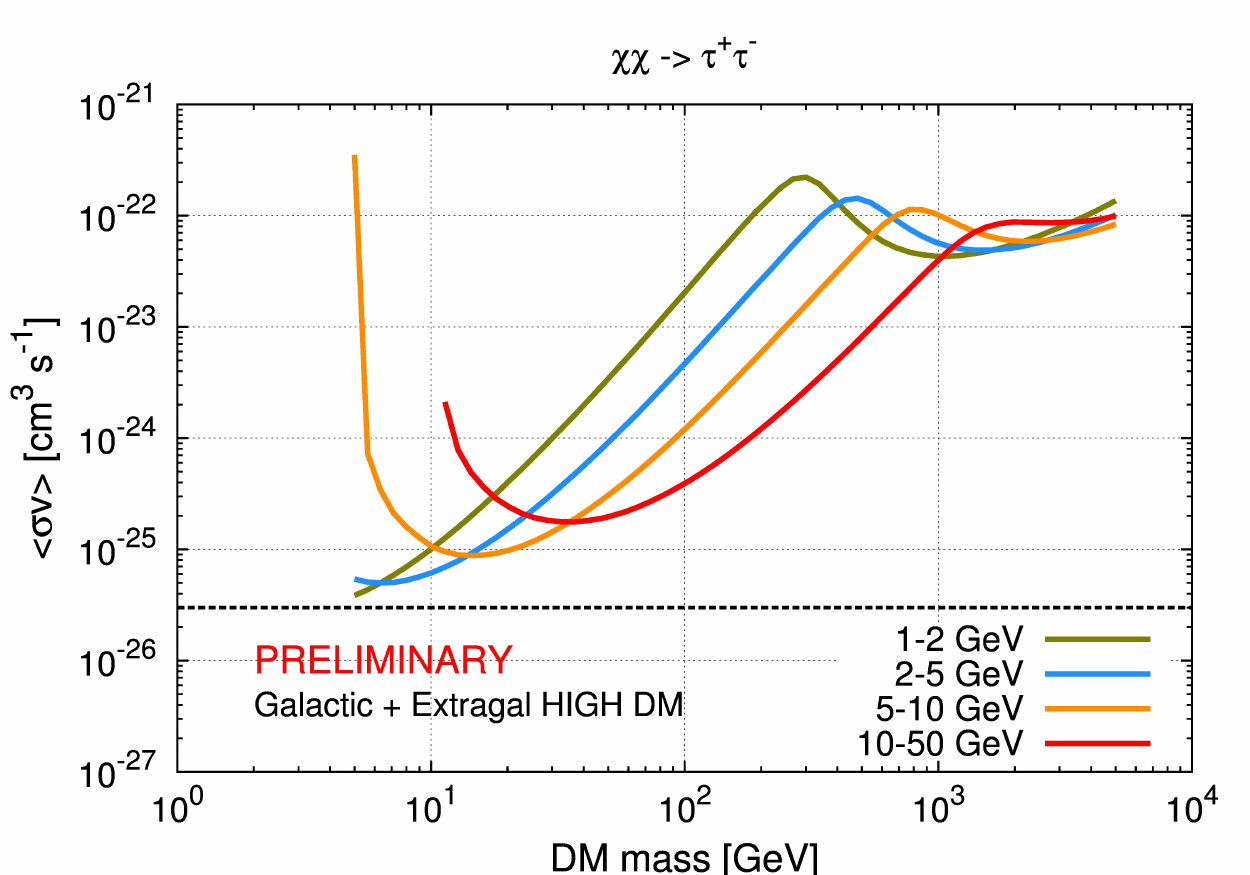}
\includegraphics[width=.45\textwidth]{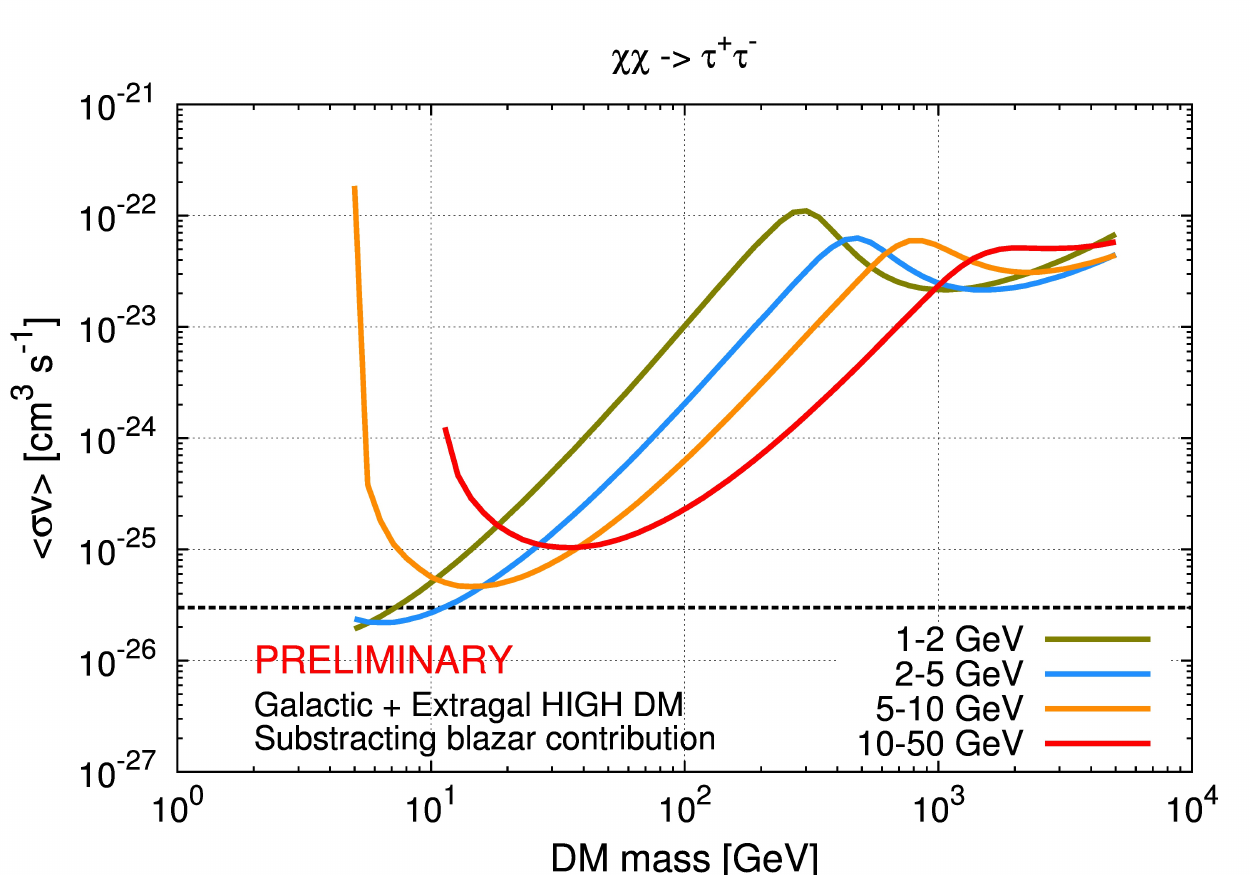}
\caption{Preliminary constraints on dark matter annihilation to $\tau^{+}\tau^{-}$ from the measured anisotropy of the IGRB~\cite{Ackermann:2012uf}.  {\it Left panel:} 95\% C.L.~upper limit on the annihilation cross section when allowing dark matter to contribute the entirety of the measured anisotropy; constraints are shown for each energy bin of the anisotropy measurement.   {\it Right panel:} 95\% C.L.~upper limit on the annihilation cross section allowing dark matter to contribute the entirety of the non-blazar anisotropy.  The limits improve by a factor of a few when accounting for the blazar contribution to the IGRB anisotropy.  From~\cite{Gomez-Vargas:2013cna}.\label{fig:dmanisolimits}}
\end{figure*}

The impact of modeling contributions from astrophysical source populations can been seen in Fig.~\ref{fig:dmanisolimits}.  The left panel shows the 95\%~C.L.~upper limit on the dark matter annihilation cross section as a function of WIMP mass, obtained with the conservative  approach of requiring only that the dark matter anisotropy does not exceed the total measured anisotropy, with the dark matter anisotropy determined as in~\cite{Fornasa:2012gu}.  The constraints in the right panel are instead obtained by requiring that the dark matter signal not exceed the 95\%~C.L.~upper limit on the non-blazar anisotropy, with the blazar anisotropy calculated in~\cite{Cuoco:2012yf} and the dark matter anisotropy again determined as in~\cite{Fornasa:2012gu}. The dark matter limits improve by a factor of a few due to modeling this one significant guaranteed component of the IGRB\@.  Tying down the anisotropy contributions of other IGRB contributors will further increase the sensitivity of dark matter searches in the IGRB\@.

\begin{figure*}[t]
\centering
\includegraphics[width=.4\textwidth]{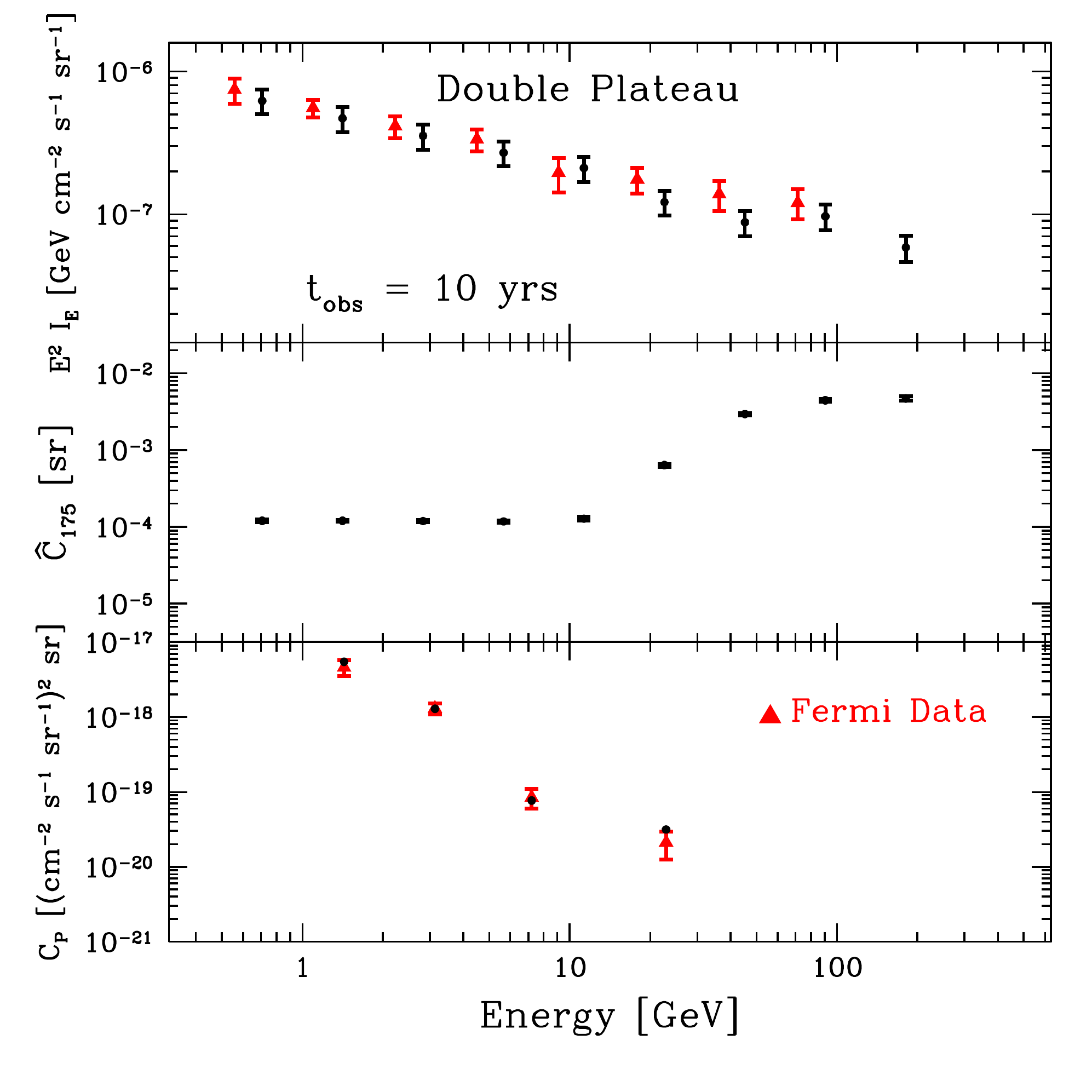}\hspace{1cm}
\includegraphics[width=.4\textwidth]{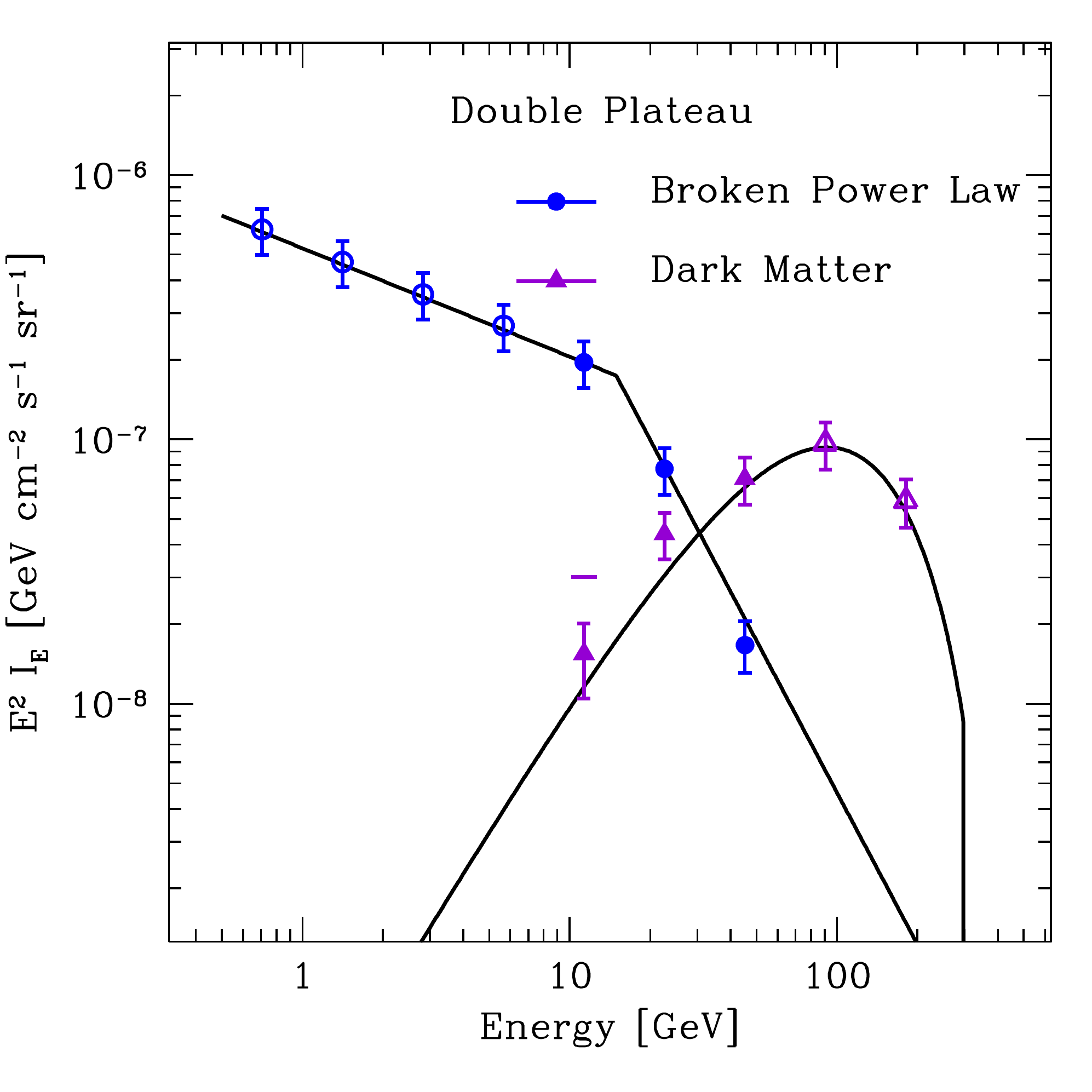}
\caption{{\it Left panel:} Simulated observations for an example two-component IGRB composition scenario assuming 10 years of all-sky observation with the \emph{Fermi} LAT (black points), consistent with current observations (red points).  {\it Right panel:} Component intensity spectra decomposed from the simulated observations shown in the left panel, using the \emph{Double Plateau} technique. From~\cite{Hensley:2012xj}.\label{fig:decon}}
\end{figure*}

As noted previously, if the dark matter signal is sufficiently subdominant that the energy spectrum of the total measured emission does not display a clear feature, identifying components of diffuse emission based on the total energy spectrum alone is challenging.  However, the distinct energy spectra of the components still provide valuable information.
New approaches to disentangle the contributions of multiple sources to diffuse emission by combining the anisotropy and 
energy spectrum information have been proposed~\cite{SiegalGaskins:2009ux,Hensley:2012xj}.  In~\cite{Hensley:2012xj} several techniques are presented which, in certain situations, allow the intensity spectra of the components of diffuse emission to be recovered using only the measured intensity energy spectrum and anisotropy energy spectrum (angular power at a fixed multipole as a function of energy).  In these situations the intensity spectrum of each component contributing to diffuse emission can be solved for analytically from the observed intensity and anisotropy energy spectra, without making \emph{a priori} assumptions about the intensity spectra or anisotropy properties of the components.  An example decomposition of a hypothetical IGRB scenario, consistent with current observations (shown in red), is illustrated in Fig.~\ref{fig:decon}.  The component spectra in the right panel are determined using only the observations shown in black in the left panel.  A two-component scenario in which the components are uncorrelated is assumed in this example.  These techniques, which rely on features observed in the anisotropy energy spectrum, can be used to model-independently identify the intensity spectra of multiple contributors to diffuse emission, and can be particularly valuable for extracting information about a subdominant contributor, such as a dark matter signal.  These decomposition techniques are also applicable to multi-wavelength data sets, allowing for more powerful analyses covering a broad energy range.

\section{Conclusions}

Indirect searches for dark matter annihilation and decay products have the unique potential to confirm the particle nature of dark matter using astrophysical observations.  With current searches already probing exciting regions of parameter space, future observations will play a critical role in advancing our knowledge of dark matter.

Some of the most promising targets of indirect searches are also the most challenging due to the presence of substantial astrophysical backgrounds.  However, making use of upcoming multi-wavelength data sets will enable far better characterization of these backgrounds, reduce uncertainties in the astrophysical emission, and thereby allow for more sensitive indirect searches.  Moreover, new approaches that incorporate angular and spectral information in previously unexplored ways can further help to meet the challenge of detecting a subdominant dark matter signal.

\begin{acknowledgments}
It is a pleasure to acknowledge the many collaborators and colleagues whose work is discussed here.  I also acknowledge support from NASA through Einstein Postdoctoral Fellowship grant PF1-120089 awarded by the Chandra X-ray Center, which is operated by the Smithsonian Astrophysical Observatory for NASA under contract NAS8-03060.
\end{acknowledgments}

\end{article}

\end{document}